\documentclass[printer]{aa}  

\usepackage{graphicx}
\usepackage{txfonts}
\defcitealias{Rodriguez19}{RLL19}

\begin{document}

%\title{The wind-wind collisions in binary LBVs}

  \title{Understanding the radio emission from $\epsilon$ Eridani}

%   \subtitle{}

   \author{Luis F. Rodr\'{\i}guez\inst{1}, Susana Lizano\inst{1},
         Jorge Cant\'o\inst{2}
          \and
         Ricardo F. Gonz\'alez\inst{1}}

  \institute{Instituto de Radioastronom\'{\i}a y Astrof\'{\i}sica,
    Universidad Nacional Aut\'onoma de M\'exico, A.P. 3-72 (Xangari) 58089
    Morelia, Michoac\'an, M\'exico\\
  \email{l.rodriguez@irya.unam.mx}
  \and
    Instituto de Astronom\'{\i}a,
   Universidad Nacional Aut\'onoma de M\'exico, A.P. 70-264, CDMX 04510, M\'exico}

   \date{Received ; accepted}

\abstract
{Some solar-type stars are known to present faint, time-variable radio continuum emission whose nature is not clearly established. We report on \textit{Jansky} Very Large Array observations of the nearby star $\epsilon$ Eridani at 10.0 and 33.0 GHz. We find that this star has flux density variations on scales down to days, hours, and minutes. On 2020 April 15 it exhibited a radio pulse at 10.0 GHz with a total duration of  about 20 minutes and a peak four times larger than the plateau of 40 $\mu$Jy present in that epoch. We were able to model the time behavior of this radio pulse in terms of the radiation from shocks ramming into the stellar wind.  Such shocks can be produced by  the wind interaction of violently expanding gas heated suddenly by energetic electrons from a stellar flare, similar to the observed solar flares.  Because of the large temperature needed in the working
surface to produce the observed emission, this has to be nonthermal. It could be gyrosynchrotron or synchrotron emission.  Unfortunately, the spectral index or polarization measurements from the radio pulse  do not have a high enough signal-to-noise ratio to allow us to determine its nature.} 
\keywords{Radio continuum emission(1340) --- K dwarf stars(879)}

 \maketitle

%________________________________________________________________

\section{Introduction}
\label{Intro}

Located at 3.22 pc, $\epsilon$ Eridani (HD 22049) is a young Sun-like
star close to our Solar System. Its spectral type is K2V, and its estimated
age is 0.8 Gyr  \citep{2004A&A...426..601D}. It has a circumstellar
debris disk with a physical extension of $\sim$ 64 AU and a radial width of $\sim$ 20 AU,
possibly the residual of the process of formation of a young planetary system
\citep{1998ApJ...506L.133G}. \cite{2009ApJ...690.1522B}  
and \cite{2011A&A...527A..57R}  have suggested that this disk consists of
different components: two warm inner belts, a cold outer belt, and an extended
halo of small grains. Observations by \cite{2016MNRAS.462.2285C}
 and \cite{2018ApJ...857..133B}  show continuum
emission from the central star at millimeter and centimeter wavelengths.

Like our Sun, $\epsilon$ Eridani exhibits an active chromosphere
 \citep[e.g.,][]{1981ApJ...246..473B}, as well as an active X-ray corona 
\cite[e.g.,] []{1981ApJ...243..234J}. In addition,  it shows a magnetic
activity cycle, which was originally reported by \cite{2000ApJ...544L.145H}. 
\cite{2013ApJ...763L..26M} report the simultaneous operation of two
magnetic activity cycles, where the amplitude of the shorter cycle of 2.95 yr
\citep[see, also,][]{2017MNRAS.471L..96J} is modulated by a longer, 12.7-year cycle.
This resembles the interaction of the 11-year solar cycle with the
quasi-biennial (2-year) variations of the Sun \citep{2010ApJ...718L..19F}.

\cite{2018ApJ...857..133B} observed radio-continuum emission associated with
$\epsilon$ Eridani with the \textit{Jansky} Very Large
Array (VLA) at four frequency bands: 2-4 GHz, 4-8 GHz, 8-12 GHz, and 12-18 GHz.  They report quasi-steady unpolarized emission in the
three high-frequency bands, and point out that this quiescent emission is
consistent with optically thick emission from the star at coronal temperatures. 
In the coronal regions of $\epsilon$ Eridani, the opacity at
frequencies higher than 2-4 GHz is dominated by gyroresonance absorption,
suggesting magnetic fields between 475 G and 2700 G in the low corona
and the production of gyroresonance emission between 4-8 GHz and 8-12 GHz.

In addition, \cite{2018ApJ...857..133B} detected a radio pulse from $\epsilon$ Eridani in the
2-4 GHz band. The detected radio pulse (a sudden brief increase in emission) had a degree of
circular polarization of up to 50\%  and lasted a few minutes.
They proposed that the origin of the radio pulse may be {nonthermal}
gyrosynchrotron emission from mildly relativistic electrons interacting with
 the coronal magnetic fields ($\sim 100$ G) of the star. 
 Alternatively, coherent cyclotron maser instability emission in the bandwidth 2-4 GHz is
also a possible mechanism for the stellar conditions, with magnetic
fields of 700-1400 kG. % Alternatively, coherent cyclotron 
%maser instability (CMI) emission is also a possible mechanism for the stellar 
%conditions with magnetic fields of 700-1400 kG. 
Nevertheless, the problem for cyclotron maser instability  
emission in coronal conditions is the escape of the radiation, since it is absorbed by 
gyroresonance processes. Alternatively, \cite{2018ApJ...857..133B} discuss the possibility that
the radio pulse could come from an as yet undetected nearby planet with a 1 kG magnetic
field, from the interaction with space weather events driven by stellar flares
and/or coronal mass ejections from $\epsilon$ Eridani.

Bastian, Benz and Gary (1998) pointed out that radio observations
were fundamental in establishing that particle acceleration in solar  flares is a two-phase process (Wild et al. 1963). There is a first phase of particle acceleration in which electrons 
reach energies of $\sim$ 100 keV, producing hard X-ray emission, microwave emission,
and type III radio bursts. A second phase occurs $\sim$ 10 min after 
the first phase, in which shock waves are produced by the initial energy released. 
These shocks then propagate into the corona. During the latter stage, electrons and ions are further accelerated by the
Fermi mechanism to energies as high as 100 MeV and 1 GeV, respectively, and type II and type
IV radio bursts occur. These bursts are thought to be responsible for geomagnetic
effects.  In addition, since their discovery,  coronal mass ejections have been recognized
as the primary drivers of interplanetary and geomagnetic perturbations (see Gosling
et al 1991, Gosling 1993). Consequently, an accepted point of view nowadays is that
interplanetary and geomagnetic disturbances are produced by two types of solar energetic
phenomena, flares and coronal mass ejections, whose relationship is still poorly understood.

Recently, Burton, MacGregor, and Osten (2022) reported the detection of three
flares from $\epsilon$ Eridani using the Atacama Large Millimeter Array (ALMA) at 1.3 mm.
They characterized the temporal, spectral, and polarization of these flare events and
suggested that their properties are similar to those of the Sun at similar wavelengths. 
Also, Loyd et al. (2022) conducted a coronal dimming analysis
at far-ultraviolet spectral lines in archival observations of $\epsilon$ Eridani  and found a prominent flare in the 2015 data.

In another work, \citet[hereafter RLL19]{Rodriguez19} reported VLA
observations of $\epsilon$ Eridani at 33 GHz. These observations had high enough angular
resolution to demonstrate the stellar origin of the radio emission since the detected emission has
an angular size of $\sim 0\rlap.{''}07$ (which corresponds to a physical extension of 0.2 au from the star), 
which is smaller than the semimajor axis ($\sim 3.4$ au) of  the exoplanet  $\epsilon$ Eridani $b$. 
In addition, \citetalias{Rodriguez19} showed that the observed centimeter emission of 
$\epsilon$ Eridani from 5 to 50 GHz is remarkably flat and could be due to optically thin
free-free emission from a steady stellar wind with a mass-loss rate of $\sim$ 6.6 $\times$
10$^{-11}$ M$_{\odot}$ yr$^{-1}$, which is 3300 times larger than the
solar value, $\dot M_\odot = 2 \times 10^{-14} M_\odot {\rm yr^{-1}}$ (Feldman et al. 1977). 
Nonetheless, other mechanisms such as coronal free-free and gyroresonance
emission could not be ruled out. The high mass-loss rate inferred by RLL19 is at odds with the mass-loss rate 
30 times the solar value estimated by Wood et al. (2002) from the interstellar hot HI that 
surrounds the star and is observed in absorption in the Ly$\alpha$ spectrum. The Ly$\alpha$ absorption comes
from the interstellar medium gas that is heated  by the interaction with the stellar wind. Wood et al. (2002) used
hydrodynamic models of astrospheres to infer the mass-loss rates from the HI absorption. 
Furthermore, Cranmer $\&$ Saar (2011) modeled the steady mass-loss rates of cool main-sequence 
stars and evolved giants. They studied magnetohydrodynamic turbulence motions from subsurface
convection zones and their dissipation and escape through the stellar wind.
They estimated an even smaller value for the mass-loss rate of
$\epsilon$ Eridani, of only 2.4 times the solar value. On the other hand, Johnstone et al. (2015) investigated
the evolution of stellar rotation and wind properties of low-mass main-sequence stars. They note that there is a large uncertainty regarding the evolution
of the stellar winds of these stars, the major problems being the lack
of observational constraints on the wind properties and the large spread in rotation rates at young ages. Thus, disagreement between the mass-loss rates obtained from observations and theoretical models of
$\epsilon$ Eridani is still an open problem.

One way to determine if the observed radio emission from $\epsilon$ Eridani 
is thermal or nonthermal is to study the
variability and/or polarization characteristics of the emission. Extremely fast variability would imply that the
source is small, and very high brightness temperatures, not achievable with thermal mechanisms,
would be required. The presence of linear or circular polarization is indicative of
synchrotron or gyrosynchrotron emission, respectively (G{\"u}del 2002). In addition, the
spectral index could help in determining the nature of the radio emission,
since negative indices  (<-0.1)  cannot be obtained with free-free emission (Rodr\'iguez et al. 1993). For this reason, we monitored the source at 10 GHz and 33 GHz, notably finding a radio pulse at 10 GHz. We investigated the
possibility that the pulse can be produced by shocks in the $\epsilon$ Eridani stellar wind. 
As discussed above, such shocks could be produced in the second phase of stellar flares,
after the magnetic energy is released.

%when gas be due to the interaction of violently expanding gas, 
%heated suddenly by energetic electrons from a stellar flare, accelerated by the
%release of magnetic energy. 

The rest of this paper is organized as follows. In Sects. 2 and 3 we discuss the observations.
In Sect. 4 we present the model of the flare--wind interaction used to interpret the radio
pulse observed at 10 GHz. In Sect. 5 we present our conclusions.

\section{Observations} 

The observations were part of our
VLA project 20A-020, made with the \textit{Karl G. Jansky} VLA of NRAO\footnote{The National 
Radio Astronomy Observatory is a facility of the National Science Foundation operated
under cooperative agreement by Associated Universities, Inc.} in its C configuration
during eight epochs in 2020 February-May. We used the X band (8-12 GHz) at six epochs and the Ka band (29-37 GHz)
at two epochs.
The flux and bandpass calibrator was J0137+3309 or J0542+4951, and
the phase calibrator was always J0339-0146.
In the X band, the digital correlator of the VLA was configured in 32 spectral windows of 128 MHz width, covering the range
8 to 12 GHz. In the Ka band,  the digital correlator of the VLA was configured in 
64 spectral windows of 128 MHz width, covering the range
29 to 37 GHz. In both bands, each spectral window is divided into 64 channels with an individual spectral resolution of 2 MHz. 
The data were calibrated in the standard manner using the Common Astronomy Software Applications (CASA) package of NRAO and
the pipeline provided for VLA\footnote{https://science.nrao.edu/facilities/vla/data-processing/pipeline} observations (McMullin et al. 2007). 

We created an image concatenating the six epochs observed at 10.0 GHz (the center of the X band) to search for sources in the field. A robust weighting of 2 (Briggs 1995) was used to optimize sensitivity in all the images discussed here. A 
total of 13 sources were detected (see Table 1). One of the radio sources corresponds to $\epsilon$ Eridani and of the remaining 12 sources, 10 had been previously detected by \citetalias{Rodriguez19}. With the exception of [RLL2019] 3, which shows an increase of a factor of 2 between the observations 
reported in  \citetalias{Rodriguez19} (taken in 2013) and those presented here, the other sources show flux densities consistent within
$\sim$25\% between the two epochs.

The region was observed at 10.0 GHz at six epochs and at 33.0 GHz (the center of the Ka band)
at two epochs, as listed in Table 2. In this table we
also give the flux densities of $\epsilon$ Eridani and of the steady source [RLL2019] 7 for each individual epoch. [RLL2019] 7  was used as a comparison source to check that the variability was not due to some systematic effect. 
While [RLL2019] 7 shows 10.0 GHz
flux densities that, within the noise, are consistent with no variability, $\epsilon$ Eridani exhibits important time variability. Two observations made on 2020 April 05 at 10.0 GHz with a time difference of 3.6 hours show a flux density increase of about 2.2. On 2020 April 15,
$\epsilon$ Eridani appeared unusually bright. A 10 GHz image of the emission at that epoch is shown in Fig. 1.
An analysis of the flux density within the time interval of the observations (approximately one hour) shows that most of the increase comes from a flare. This flare had a duration of only about 20 minutes and shows the characteristic rapid rise and slower fall of flares (see Fig. 2). We tried to estimate the spectral index of the emission 
during the flare by making images at
9.0 and 11.0 GHz. The flux densities obtained, 155$ \pm$17 and 175$ \pm$24 $\mu$Jy respectively, imply a spectral index of 0.6$\pm$0.9, which, unfortunately, does not have a high enough signal-to-noise ratio to discriminate between possible emission mechanisms.
The poor signal-to-noise ratio comes from the fact that the two frequencies used are close to each other.

At 33 GHz, $\epsilon$ Eridani was detected only in the observations of 2020 May 08. Given the small primary beam and
lower sensitivity at this
frequency, this was the only source detected. Between the two epochs observed at 33 GHz, with a separation of 10 days, the flux density decreased by a factor of at least $\sim$2. Again, we tried to estimate the spectral index of the emission 
on 2020 May 08 by making images at
31.0 and 35.0 GHz. The flux densities obtained, 62$ \pm$15 and 61$ \pm$20 $\mu$Jy respectively, imply a spectral index of -0.1$\pm$3.4, which, again, does not have a high enough signal-to-noise ratio to discriminate between possible emission mechanisms.
Finally, the last column of Table 2 gives the 4$\sigma$ upper limit to the circular polarization of $\epsilon$ Eridani. No circular polarization was detected at any epoch.

\section{Individual sources}

\subsection{Sources 2 and 13}

As noted by  \citetalias{Rodriguez19}, source 11 of Table 1 is associated with a millimeter source detected by 
\cite{2016MNRAS.462.2285C}.
In the $7\rlap.'68 \times 7\rlap.'68$ region we imaged at 10.0 GHz, a total of 13 radio sources were detected (Table 1). We wanted to include optical counterparts other than $\epsilon$ Eri and thus added the \sl Gaia \rm Data Release 3 sources from the same region of the sky to our sample
(Gaia collaboration et al. 2016; 2022), for a total of 64 sources. Remarkably, radio sources
2 and 13 coincide within $0\rlap.''6$ with the sources  \textit{Gaia} DR3 5164707626664526592 and 5164696081792414464, respectively.
We speculate that these two \textit{Gaia} sources could be background galaxies. According to the NASA/IPAC Extragalactic Database (NED), only source 2 is an
infrared source. However, there are no other data, and the nature of the
source (galactic or extragalactic) is not determined in the NED.

\subsection{Source 11}

This source was detected as a single source in the 2013 observations presented by \citetalias{Rodriguez19}, which had an angular resolution of
$\sim 6{''}$. The higher-angular-resolution observations presented here (Fig. 3) show it is actually a double
source, probably a radio galaxy. 

\section{The radio behavior of $\epsilon$ Eridani}

\subsection{Comparison with the Sun}

At 10.0 GHz, the quiet Sun (at sunspot minimum) has a flux density of $\simeq 2.8 \times 10^{6}$ Jy as measured from the Earth (Giersch \& Kennewell 2022). A typical plateau flux density for 
$\epsilon$ Eridani at 10.0 GHz is $\simeq$ 40 $\mu$Jy (this paper). If it were located at 1 au, we would measure a flux density of $\simeq 1.8 \times 10^{7}$ Jy, about eight times larger than that of the Sun.

With respect to flare activity, Nita et al. (2002) investigated the peak flux distribution of 40 years of solar radio burst data. 
They find that in the 8.4–11.8 GHz band during solar maximum, one can expect one flare above $8 \times 10^7$ Jy about every
30 days. The total peak flux density of 190 $\mu$Jy of the $\epsilon$ Eridani flare reported here
would become $8.4 \times 10^7$ Jy at 1 au, comparable to the brightest solar flares. On the other hand, 
Wang et al. (2020) studied 86 solar events at 9.4 GHz between 2000 and 2010. They find that the peak flux densities are distributed along two orders of magnitude, from about $2  \times 10^5$ to about $2  \times 10^7$ Jy, but the relatively small number of
flares observed limits their statistics.

\subsection{Interpretation of the observed radio pulse at 10 GHz} \label{sec:interpretation}

Unfortunately, it was not possible to determine the spectral index of the emission during the radio pulse at 10 GHz
accurately enough to determine if the origin of the emission is thermal or nonthermal. As discussed in Sect. 2, the derived value of
the spectral index is  0.6$\pm$0.9. 

The observed pulse at 10 GHz could be produced by shocks in the ionized stellar
wind of $\epsilon$ Eridani (see \citetalias{Rodriguez19}). As mentioned above, shocks  can occur from the
interaction of gas  heated by energetic electrons from a stellar  flare  with the steady stellar wind of the star. 
When the heated gas collides with the stellar wind,  a working surface (WS), bounded by two shock
fronts, will be formed as a fast flow overtakes the previously ejected slow wind. In this section we model the
observed radio pulse at 10 GHz of $\epsilon$ Eridani as produced by shocks in the stellar wind inside a cone
with solid angle $\Omega$. 

%We will use the model of the dynamics and luminosity of the WS by  \cite{2000MNRAS.313..656C}, 
%for highly supersonic flows, that is based on mass and momentum conservation. 

We assumed that there is an increase in the flow velocity at the  base of the wind, that is,
during the violent expansion event the ejection velocity drastically increases with respect
to the stellar wind value.  We also allowed the expanding gas to have a mass-loss rate different 
from that of the stellar wind. This type of variability in the flow parameters has previously been studied
by \cite{2022MNRAS.509.1892M} in the case of the Sun. They used the general model of the WS dynamics from  \cite{2000MNRAS.313..656C} for highly supersonic flows, which is based on mass and momentum conservation. 
%\sout{using the analytic method developed by \cite{2000MNRAS.313..656C} to solve the
%equations for highly supersonic flows, based on mass and momentum conservation.} 

The  \cite{2022MNRAS.509.1892M} model gives an analytic solution for the dynamical evolution of the WS. In this model, 
the wind is ejected with a constant velocity $v_1 = v_{w}$ and an isotropic mass-loss rate $\dot{M_w}$. At an injection time $\tau = 0$, one considers mass injection within a solid angle $\Omega$, with an increased velocity $v_{2}\, = a\,v_1$, with 
$a >1$, during an interval of time $\Delta{\tau}$, and a change in the mass-loss rate per unit solid angle from the wind mass-loss rate $\dot m_1 = \dot M_w/ 4 \pi$ to  $\dot m_2 = b \dot m_1$.
 In this case, the WS forms instantaneously and has an initial phase where it 
 moves at a constant speed, $v_{\rm WS} = \sigma v_w$, where $\sigma = a^{1/2} ( 1 + a^{1/2} b^{1/2})/(a^{1/2} + b^{1/2})$. 
Consequently, the position of the WS during this initial phase is given by $r_{\rm WS}= R_{*} + \sigma v_{w} t$, where $R_{*}$ is the
injection radius. Once all the heated gas has entered the WS,
the inner shock disappears, giving rise to a decelerating phase. This occurs at a critical
time $t_{c}= a \, \Delta{\tau} / (a - \sigma)$. For later times, the WS decelerates with a velocity and position given 
by Eqs. (2) and (3) of \cite{2022MNRAS.509.1892M}.
The WS moves away from the source and decelerates,
asymptotically approaching the wind velocity, $v_{w}$. During its evolution, the WS will lose
energy via radiation. From Eqs.  (4), (5), and (19) of   \cite{2000MNRAS.313..656C}, the
bolometric luminosity of the WS (within the solid angle $\Omega$) can be written
as
\begin{eqnarray}
L_{\rm WS} (t) =  \left({\Omega \over 4 \pi} \right)  {1\over 2} \dot M_w v_w^2 
\left\{ 
{b\over a} \left(a - {v_{\rm WS} \over v_{w} } \right)^3 
+
 \left({v_{\rm WS} \over v_{w} }-1 \right)^3 
  \right\}
 \label{eq:Lbol}
\end{eqnarray}
for $t \le t_c$, when the WS is bounded by two shocks, and
\begin{eqnarray}
L_{\rm WS} (t) =  \left({\Omega \over 4 \pi} \right)  {1\over 2} \dot M_w v_w^2 
 \left({v_{\rm WS} \over v_{w} }-1 \right)^3 
 \label{eq:Lbol2}
\end{eqnarray}
for $t > t_c$, when the internal shock disappears and the WS decelerates.

For simplicity, we assumed that a constant fraction, $\epsilon$, of the WS bolometric luminosity  is radiated at 10 GHz. Thus, 
the 10 GHz flux  is given by 
\begin{equation}
F_{\rm 10 \, GHz}(t)  = \epsilon F_{\rm WS} \equiv \epsilon { L_{\rm WS}(t) \over 4 \pi D^2 },
\label{eq:Flux}
\end{equation}
where  $F_{\rm WS} $ is the bolometric flux, and  $D=3.22$ pc is the distance to $\epsilon$ Eridani.

To correct the ten data points of the 2020 April 15 observations, we fitted and subtracted the slope of a baseline in time obtained
from a linear fit to the first four points. The resulting flux densities are given in Table 3. In Fig. \ref{Fig:flare} we show these
data points as well as a model, as a solid red line. This model gives the flux from a 
WS produced in a  wind with mass-loss rate $\dot M_w = 
6.6 \times 10^{-11} M_\odot {\rm yr^{-1}} $ and velocity $v_w= 650 \, {\rm km s^{-1}}$ (taken from the model of RLL2019 for the steady wind of $\epsilon$ Eridani),  a velocity increase $a= 3.3$, a mass-loss rate parameter $b=2.6$, a time interval of injection of high velocity gas $\Delta \tau= 3.0$ min, and a solid angle $\Omega = 9.5 \times 10^{-2}$ str (corresponding to a half opening angle $\theta=10^\circ$). 
The emission of the WS is superimposed on a steady wind emission of 40 $\mu$Jy. 
The WS has initial velocity $v_{WS} =  1300  \, {\rm km s^{-1}}$. The inner shock disappears at the critical time $t_c= 8.2$ min, and then the WS starts to decelerate and eventually reaches the wind velocity. The deceleration starts at  a distance $R_c=1.7 \, R_\odot$ from the center of the star. We note that the duration of the observed radio pulse is longer than the model time interval of ejection of fast material ($\Delta \tau$).

The observed maximum flux of the radio pulse is $150 \, \mu Jy$, emitted on top of a steady wind emission of $40  \, \mu Jy$. %, such that the total observed flux is $190 \, \mu Jy$.  
For the shock model discussed above, the maximum of the total bolometric flux  is $F_{\rm max} = 1.4 \times 10^{13} Jy$  (obtained by substituting $v_{\rm WS}/ v_w = \sigma$ in Eqs. \ref{eq:Lbol} and \ref{eq:Flux}). To observe $150 \, \mu Jy$ at the peak of the 10 GHz pulse, one needs a fraction, $\epsilon$, of the total energy emitted at this frequency to be
 $\epsilon  = 150 \, \mu Jy/ F_{\rm max} = 1.0 \times 10^{-17}$.
 On the other hand, as discussed in the Introduction, Wood et al. (2002) estimated a mass-loss rate for $\epsilon$ Eridani   of 30 times the solar value, that is to say, 110 times smaller than the value of RLL19. Assuming the  same wind speed (which is the escape speed from the 
 surface of the star), the shock would have
a maximum flux, $F_{\rm max}$, 110 times smaller because the shock luminosity is proportional to $\dot M_w$
(Eq. \ref{eq:Lbol}). In this case,
the fraction of the total energy emitted at 10 GHz would be 110 times higher, namely
 $\epsilon  =  1.2 \times 10^{-15}$.
 
In this simple model, we do not solve for the shock microphysics. Nevertheless, it can be used to obtain further insights into the 
origin of the emission. First, the estimated size, $l$, of the emitting region at the distance 
$R_c=1.7 \, R_\odot$ with opening angle  $\theta=10^\circ$ is $l \sim 2.1 \times 10^{10} {\rm cm}$. Thus, at the distance of $\epsilon$ Eridani, the WS would need to have a brightness temperature $T_b \sim 3.6 \times 10^6$ K  to produce the observed peak flux of 150 $\mu Jy$ at 10 GHz.

If the emission is thermal, one can calculate the fraction of energy emitted by a black body (BB) in the range 8-12 GHz. 
The integral over the whole spectrum, $B_\nu(T),$  is 
$ I_{\rm total} = \int_0^\infty B_\nu d \nu = {\sigma_B T^4 / \pi} \quad {\rm erg \, s^{-1} cm^{-2}  str^{-1}} ,$ where $\sigma_B$ is the Stefan-Boltzmann constant. In  the Rayleigh-Jeans approximation, the integral over the VLA bandwidth at 10 GHz is 
$ I_{\rm 10 GHz } = \int_{\rm 8 GHz}^{\rm 12 GHz}  B_\nu d \nu  \sim 
\left({2 k T \over 3 c^2} \right) (12^3 - 8^3) \times 10^{27}  \times 
 {\rm erg \, s^{-1} cm^{-2}  str^{-1}}.$ Then, the fraction of the BB energy that is radiated in the 10 GHz band is
$ \epsilon_{BB} = {I_{\rm 10 GHz } / I_{\rm total}} = 1.4 \times 10^{-22} \left( { T /3.6 \times 10^6 K } \right)^{-3} . $
The fraction of thermal BB energy emitted at 10 GHz is thus too small compared to the values calculated above and so the emission of the WS  cannot be thermal emission. Therefore, the emission produced by the WS needs to be nonthermal, for example synchrotron emission by relativistic electrons accelerated in the shocks via the Fermi mechanism.
In the Sun after a solar flare, electrons are Fermi-accelerated in shocks
up to energies of 100 Mev (e.g., Bastian et al. 1998).

In summary, internal shocks produced in the wind, expected from gas heated by stellar flares, can produce the temporal variation in the observed 10 GHz pulse in $\epsilon$ Eridani. The emission of the WS has to be nonthermal.
Further monitoring of this source is needed to obtain the spectral index or polarization
measurements of the radio emission and thus further constrain the emission mechanism.
The nonthermal emission could be due to gyrosynchrotron from a flare due to magnetic reconnection 
in the low corona (Nita et al. 2002) or to synchrotron emission from relativistic electrons accelerated in shocks, as in the scenario discussed above. Further observations are required to determine the polarization state and spectral index of the emission. 
A positive spectral index would indicate gyrosynchrotron or optically thick synchrotron emission. 
In addition, gyrosynchrotron would produce circularly polarized emission, while one expects linear polarization in the case of synchrotron emission (G{\"u}del 2002). In particular, the observed radiation is entirely consistent with nonthermal gyrosynchrotron emission from a solar-like flare in the low corona. For example, for a source radius of 5\% of the stellar radius, the brightness temperature 
would be a few 
times $10^8$ K, consistent with nonthermal gyrosynchrotron emission. This interpretation is also supported by the fact that at lower
frequencies Bastian et al. (2018) doubtless detected a gyrosynchrotron pulse.
 
\section{Conclusions}

Our main conclusions can be summarized as follows:

%\begin{itemize}

 We present sensitive, high-angular-resolution VLA observations of the nearby (3.22 pc) Sun-like star epsilon Eridani
 at 10.0 (six epochs) and 33.0 GHz (two epochs). While the other radio sources in the field are steady, $\epsilon$ Eridani shows
 flux density variations down to scales of days, hours, and even minutes. The emission at 10 GHz, excluding the radio pulse 
 observed  on 2020 April 15,  varied by a factor of $\sim$ 3 during the observation campaign.

The 2020 April 15 radio pulse of $\epsilon$ Eridani lasted for about 20 minutes. At the
 peak of the pulse, the flux density was $\sim$150 $\mu$Jy, a factor of \textasciitilde4 above the plateau of 
 $\sim$40 $\mu$Jy present at that epoch.

We successfully modeled the energy and temporal variation in the 10 GHz radio pulse in terms of the emission from shocks in the wind 
of $\epsilon$ Eridani. These shocks have enough energy to produce the observed flux. 
The model does not solve for the microphysics of the shock emission. Nevertheless, the brightness temperature required 
to produce the observed flux implies that the emission has to be nonthermal.

Unfortunately, neither the spectral index nor the polarization measurements of the radio flare emission are stringent enough to
 determine the nature of the nonthermal radio emission, either gyrosynchrotron from stellar flares or synchrotron emission
 from relativistic electrons accelerated in shocks. However, the observed flare
 radiation is consistent with nonthermal gyrosynchrotron emission from a solar-like flare in the low corona. Ultra-sensitive observations such as those that will be provided by the
 Square Kilometer Array and the Next Generation VLA may be required to firmly establish the nature of the
 radio emission of $\epsilon$ Eridani and similar stars.
 
 %\end{itemize}

\bigskip
\noindent {\it Acknowledgements:}  This work was supported by grants PAPIIT-UNAM  IN103921, IG100422, IN103023.
This research has made use of the NASA/IPAC Extragalactic Database (NED), which is funded by the National Aeronautics and 
Space Administration and operated by the California Institute of Technology. This work has made use of data from the 
European Space Agency (ESA) mission
{\it Gaia} (\url{https://www.cosmos.esa.int/gaia}), processed by the {\it Gaia}
Data Processing and Analysis Consortium (DPAC,
\url{https://www.cosmos.esa.int/web/gaia/dpac/consortium}). Funding for the DPAC
has been provided by national institutions, in particular the institutions
participating in the {\it Gaia} Multilateral Agreement.

We thank the referee for very useful comments that improved the interpretation of the data.

{}{}

\begin{table*}
%\tablenum{1}
\caption{10 GHz sources in the field of $\epsilon$ Eridani.\label{tab:field}}
\label{table:1}      
\centering
\begin{tabular}{c c c c c}
%\tablewidth{0pt}
%\tablehead{
\hline\hline
 & RA(J2000) & DEC(J2000) & Flux Density &  \\
Number & $03^h$ & $-09^ \circ$ & ($\mu$Jy) & Counterparts  \\
%}
%\decimalcolnumbers
%\startdata
\hline
1 & $32^m~44\rlap.^s343$  & $29'~21\rlap.{''}14$ & $180\pm29$ &  --- \\
2 & $32^m~47\rlap.^s004$  & $28'~23\rlap.{''}59$ & $127\pm12$ &  [RLL2019] 3  \\
3 & $32^m~47\rlap.^s783$  & $25'~43\rlap.{''}28$ & $70\pm15$ &  --- \\
4 & $32^m~48\rlap.^s852$  & $27'~10\rlap.{''}53$ & $41\pm7$ &  [RLL2019] 4 \\
5 & $32^m~49\rlap.^s351$  & $28'~16\rlap.{''}49$ & $250\pm9$ &  [RLL2019] 5 \\
6 & $32^m~54\rlap.^s498$  & $27'~29\rlap.{''}37$ & $46\pm5$ &  $\epsilon$ Eridani \\
7 & $32^m~55\rlap.^s149$  & $28'~23\rlap.{''}02$ & $178\pm7$ &  [RLL2019] 7 \\
8 & $32^m~56\rlap.^s148$  & $25'~29\rlap.{''}96$ & $63\pm6$ &  [RLL2019] 8 \\
9 & $32^m~58\rlap.^s011$  & $24'~13\rlap.{''}62$ & $352\pm33$ &  [RLL2019] 9 \\
10 & $32^m~58\rlap.^s144$  & $25'~05\rlap.{''}95$ & $1127\pm25$ &  [RLL2019] 10 \\
11 & $32^m~59\rlap.^s068$  & $28'~29\rlap.{''}40$ & $72\pm11$ &  [RLL2019] 11 \\
12 & $32^m~59\rlap.^s899$  & $27'~52\rlap.{''}38$ & $40\pm5$ &  [RLL2019] 13 \\
13 & $33^m~08\rlap.^s730$  & $28'~10\rlap.{''}61$ & $359\pm53$ &  [RLL2019] 14 \\
\hline
\end{tabular}
\tablefoot{The positions and flux densities are from an image made concatenating the data from the
six epochs observed at 10.0 GHz. The flux densities are corrected for the primary beam response.}
\end{table*}

\begin{table*}
\caption{Flux density of $\epsilon$ Eridani and [RLL2019] 7 as a function of time (averaged over each epoch of observation). \label{tab:time}}
\centering
\begin{tabular}{ccccc}
\hline\hline
 & Frequency &  \multicolumn{2}{c}{Flux Density($\mu$Jy) }  &  Circular \\
Epoch & (GHz) & $\epsilon$ Eridani & [RLL2019] 7  & Polarization \\
\hline
2020 Feb 11 & 10.0 & $37.5\pm5.1$  & $216\pm12$ &$\leq$44\% \\
2020 Feb 19 & 10.0 & $46.8\pm3.3$ & $193\pm9$ & $\leq$26\% \\
2020 Apr 05 & 10.0 & $17.3\pm3.4$  & $170\pm12$ & $\leq$86\% \\
2020 Apr 05 & 10.0 &  $37.5\pm3.1$ & $193\pm11$ & $\leq$40\% \\
2020 Apr 15 & 10.0 & $79.6\pm3.1$ & $191\pm9$ & $\leq$19\% \\
2020 Apr 17 & 10.0 & $29.9\pm3.9$ & $162\pm13$ & $\leq$65\% \\
2020 May 08 & 33.0 & $62\pm12$ &  --- & $\leq$58\% \\
2020 May 18 & 33.0 & $ \leq$32 &  --- &  --- \\
\hline
\end{tabular}
\tablefoot{The last column is the 4$\sigma$ upper limit to the circular polarization of $\epsilon$ Eridani.
The two observations from 2020 April 05 were made with a time difference of 3.6 hours. The values given for
all epochs are from average images made using all the data from that day. During the flare of 2020 April 15, 
the 10.0 GHz flux density reached a value
of 165$\pm$12 $\mu$Jy and the upper limit for the circular polarization was $\leq$17\%. The value of the
33.0 GHz flux density quoted for 
2020 May 18 is a 4$\sigma$ upper limit.}

\end{table*}

\begin{table*}
\caption{Flux density of $\epsilon$ Eridani at 10 GHz on 2020 April 15.\label{tab:time2}}
\centering
\begin{tabular}{cc}
\hline
Hour & Flux Density \\ 
 (UTC) & ($\mu$Jy)  \\ 
 \hline\hline
21.369  & 44.8$\pm$8.4 \\
21.431  & 33.4$\pm$7.4 \\
21.513  & 39.3$\pm$9.5 \\ 
21.576   & 42.5$\pm$6.0 \\
21.657 & 150.1$\pm$8.6 \\
21.720 & 190.2$\pm$11.0 \\
21.801  & 107.7$\pm$9.0 \\
21.864  & 64.2$\pm$ 9.0 \\
21.946  & 69.2$\pm$8.6 \\
22.009  & 62.5$\pm$7.9 \\
\hline
\end{tabular}
\tablefoot{The flux densities have been corrected by removing the slope of a linear baseline in time.}

\end{table*}

\begin{figure}
\centering
\includegraphics[angle=0,width=0.5\textwidth]{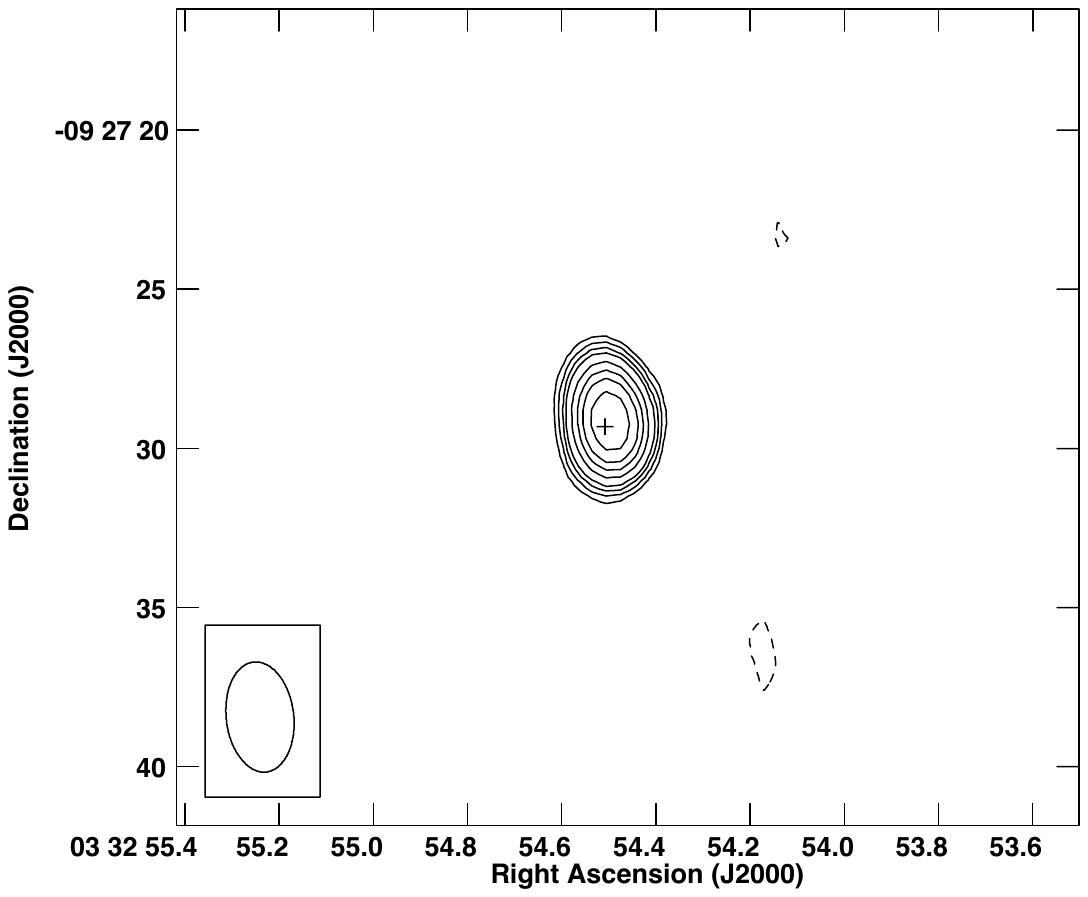}
%\vskip-2.0cm
\caption{10.0 GHz image of $\epsilon$ Eridani for the epoch 2020 April 15. Contours are -4, -3, 3, 4, 5, 6, 8, 10, 12, 
and 15 times 4.0 $\mu$Jy beam${-1}$, the rms noise in this region of the image. 
The synthesized beam ($3\rlap.{''}48 \times 2\rlap.{''}12$; PA =  6.7$^\circ$) is shown in the bottom-left corner.
The cross marks the position of the optical star from \textit{Gaia} Early Data Release 3, corrected for the proper motions of the source (Gaia Collaboration 2020).
}
\label{fig:fig1}
\end{figure}

\begin{figure}
\centering
\includegraphics[angle=0,width=0.5\textwidth]{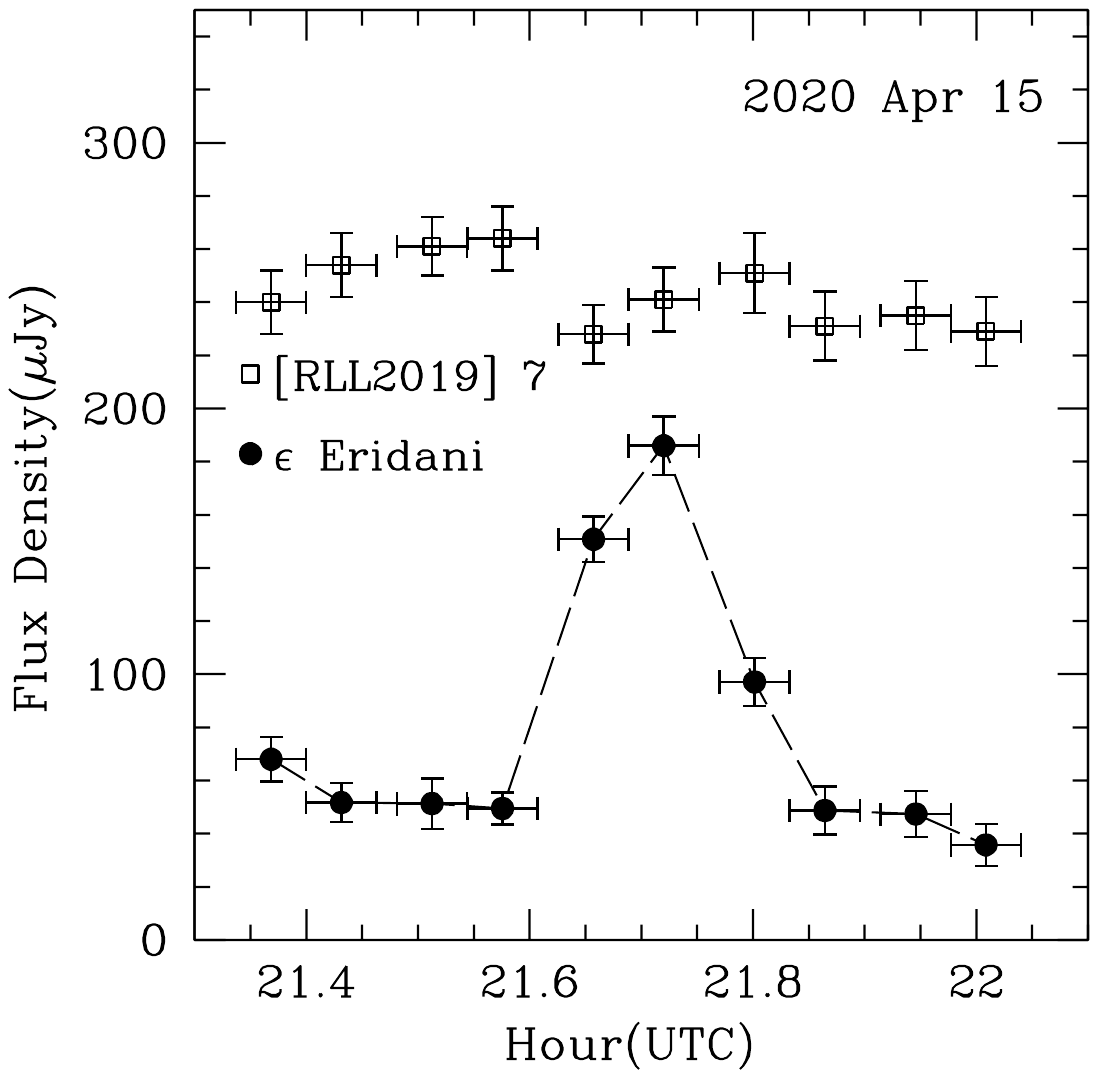}
%\plotone{epseri.pdf}
%\vskip-2.0cm
\caption{10.0 GHz flux densities for $\epsilon$ Eridani (solid circles) and [RLL2019] 7 (hollow squares)
during the observations of 2020 April 15. The hour is given in Coordinated Universal Time (UTC).
The horizontal bars indicate the duration of the scan, while the vertical bars indicate
the noise of the measurement. The dashed line joins the data points for $\epsilon$ Eridani. We have added 80 $\mu$Jy to
the flux densities of [RLL2019] 7 to better separate the data points of the two sources.
}
\label{fig:fig2}
\end{figure}

\begin{figure}
\centering
\includegraphics[angle=0,width=0.5\textwidth]{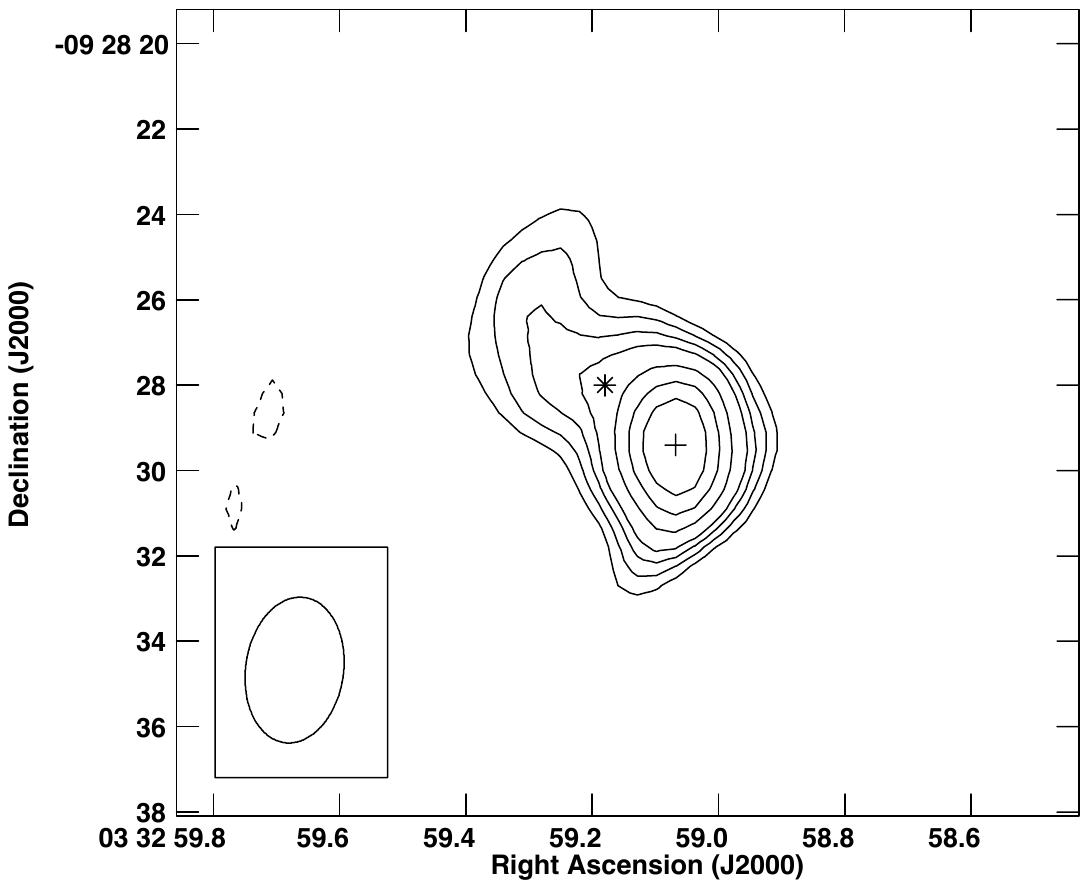}
%\plotone{EPSEXVLA11.pdf}
%\vskip-2.0cm
\caption{10.0 GHz image of the VLA11 region from all the data obtained in 2020. 
Contours are -4, -3, 3, 4, 5, 6, 8, 10, and 12  times 2.4 $\mu$Jy beam${-1}$, the rms noise in this region of the image. 
The synthesized beam ($3\rlap.{''}43 \times 2\rlap.{''}30$; PA =  -7.9$^\circ$) is shown in the bottom-left corner.
The cross marks the peak position of the radio emission in the observations reported here, while the
asterisk marks the peak position of the 10.0 GHz emission in the lower-angular-resolution
2013 observations discussed by RLL2019.} 
\label{fig:fig3}
\end{figure}

\begin{figure} 
\centering
\includegraphics[angle=0,width=0.5\textwidth]{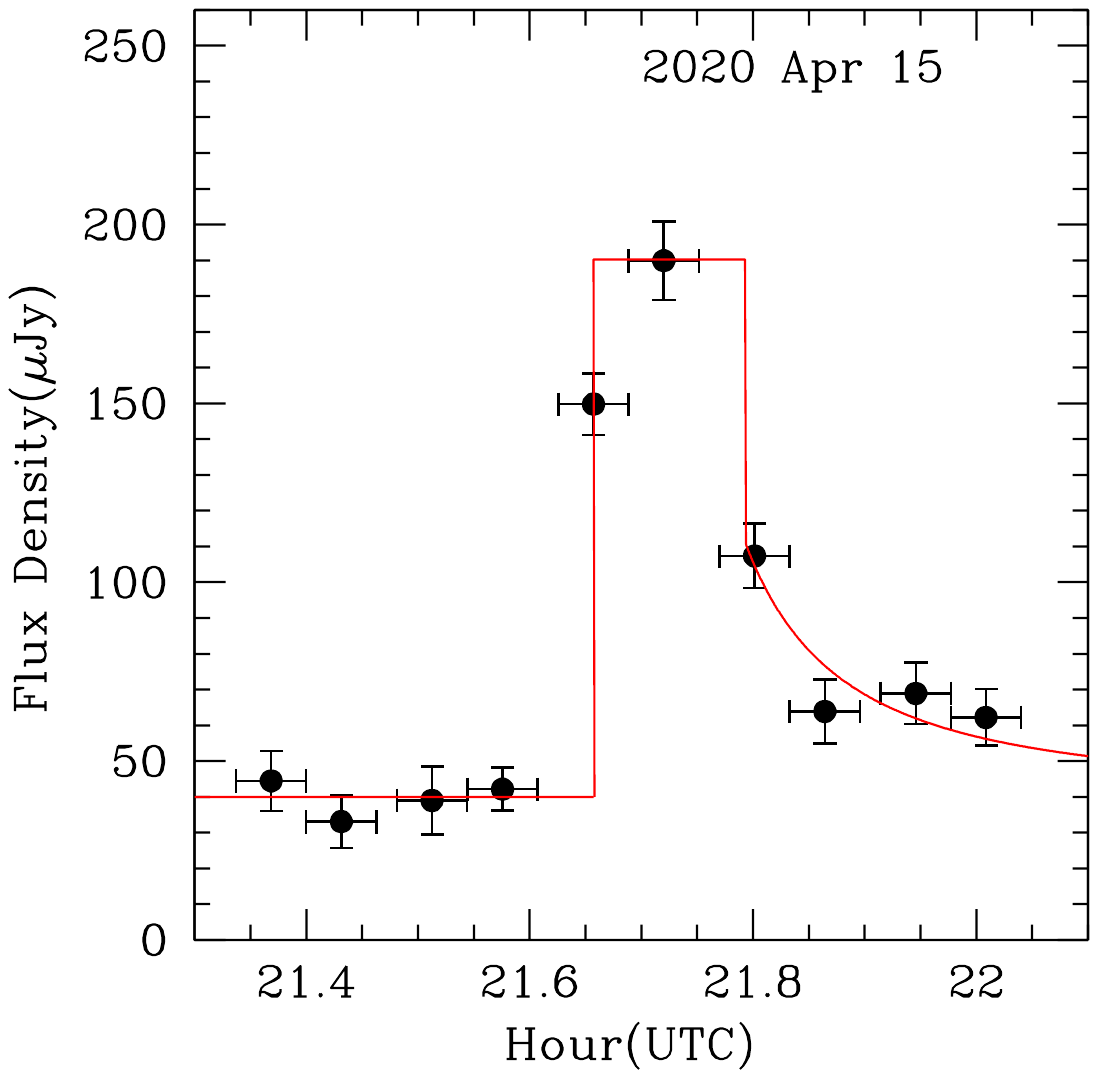}
%\vskip-2.0cm
\caption{
Model of the radio pulse as the emission of a WS formed in the stellar wind (solid red  line) superimposed on a steady emission of 40 $\mu$Jy. The data points are those given in Table 3.
}
\label{Fig:flare}
\end{figure}


\bigskip

\begin{thebibliography}{}

\bibitem[Backman et al.(2009)]{2009ApJ...690.1522B} Backman, D., Marengo, M., Stapelfeldt, K., et al.\ 2009, \apj, 690, 1522. doi:10.1088/0004-637X/690/2/1522

\bibitem[Baliunas et al.(1981)]{1981ApJ...246..473B} Baliunas, S.~L., Hartmann, L., Vaughan, A.~H., et al.\ 1981, \apj, 246, 473. doi:10.1086/158946

\bibitem[Bastian et al.(1998)]{1998ARA&A..36..131B} Bastian, T.~S., Benz, A.~O., \& Gary, D.~E.\ 1998, \araa, 36, 131. doi:10.1146/annurev.astro.36.1.131

\bibitem[Bastian et al.(2018)]{2018ApJ...857..133B} Bastian, T.~S., Villadsen, J., Maps, A., et al.\ 2018, \apj, 857, 133. doi:10.3847/1538-4357/aab3cb

\bibitem[Benedict et al.(2006)]{2006AJ....132.2206B} Benedict, G.~F., McArthur, B.~E., Gatewood, G., et al.\ 2006, \aj, 132, 2206. doi:10.1086/508323

\bibitem[Briggs(1995)]{1995AAS...18711202B} Briggs, D.~S.\ 1995, Bulletin of the American Astronomical Society, Vol. 27, p.1444.

\bibitem[Burton et al.(2022)]{2022ApJ...939L...6B} Burton, K., MacGregor, M.~A., \& Osten, R.~A.\ 2022, \apjl, 939, L6. doi:10.3847/2041-8213/ac9973

\bibitem[Cant{\'o} et al.(2000)]{2000MNRAS.313..656C} Cant{\'o}, J., Raga, A.~C., \& D'Alessio, P.\ 2000, \mnras, 313, 656. doi:10.1046/j.1365-8711.2000.03244.x

\bibitem[Chavez-Dagostino et al.(2016)]{2016MNRAS.462.2285C} Chavez-Dagostino, M., Bertone, E., Cruz-Saenz de Miera, F., et al.\ 2016, \mnras, 462, 2285. doi:10.1093/mnras/stw1363

\bibitem[Cranmer \& Saar(2011)]{2011ApJ...741...54C} Cranmer, S.~R. \& Saar, S.~H.\ 2011, \apj, 741, 54. doi:10.1088/0004-637X/741/1/54

\bibitem[Di Folco et al.(2004)]{2004A&A...426..601D} Di Folco, E., Th{\'e}venin, F., Kervella, P., et al.\ 2004, \aap, 426, 601. doi:10.1051/0004-6361:20047189

\bibitem[Feldman et al.(1977)]{1977soiv.conf..351F} Feldman, W.~C., Asbridge, J.~R., Bame, S.~J., et al.\ 1977, The Solar Output and its Variation, 351

\bibitem[Fletcher et al.(2010)]{2010ApJ...718L..19F} Fletcher, S.~T., Broomhall, A.-M., Salabert, D., et al.\ 2010, \apjl, 718, L19. doi:10.1088/2041-8205/718/1/L19

\bibitem[Gaia Collaboration et al.(2016)]{2016A&A...595A...1G} Gaia Collaboration, Prusti, T., de Bruijne, J.~H.~J., et al.\ 2016, \aap, 595, A1. doi:10.1051/0004-6361/201629272

\bibitem[Gaia Collaboration(2020)]{2020yCat.1350....0G} Gaia Collaboration\ 2020, VizieR Online Data Catalog, I/350

\bibitem[Gaia Collaboration et al.(2022)]{2022arXiv220800211G} Gaia Collaboration, Vallenari, A., Brown, A.~G.~A., et al.\ 2022, arXiv:2208.00211. doi:10.48550/arXiv.2208.00211

\bibitem[Giersch \& Kennewell(2022)]{2022RaSc...5707456G} Giersch, O. \& Kennewell, J.\ 2022, Radio Science, 57, e2022RS007456. doi:10.1029/2022RS007456

\bibitem[Gosling et al.(1991)]{1991JGR....96.7831G} Gosling, J.~T., McComas, D.~J., Phillips, J.~L., et al.\ 1991, \jgr, 96, 7831. doi:10.1029/91JA00316

\bibitem[Gosling(1993)]{1993JGR....9818937G} Gosling, J.~T.\ 1993, \jgr, 98, 18937. doi:10.1029/93JA01896

\bibitem[Greaves et al.(1998)]{1998ApJ...506L.133G} Greaves, J.~S., Holland, W.~S., Moriarty-Schieven, G., et al.\ 1998, \apjl, 506, L133. doi:10.1086/311652

\bibitem[G{\"u}del(2002)]{2002ARA&A..40..217G} G{\"u}del, M.\ 2002, \araa, 40, 217. doi:10.1146/annurev.astro.40.060401.093806

\bibitem[Hatzes et al.(2000)]{2000ApJ...544L.145H} Hatzes, A.~P., Cochran, W.~D., McArthur, B., et al.\ 2000, \apjl, 544, L145. doi:10.1086/317319

\bibitem[Jeffers et al.(2017)]{2017MNRAS.471L..96J} Jeffers, S.~V., Boro Saikia, S., Barnes, J.~R., et al.\ 2017, \mnras, 471, L96. doi:10.1093/mnrasl/slx097

\bibitem[Johnson(1981)]{1981ApJ...243..234J} Johnson, H.~M.\ 1981, \apj, 243, 234. doi:10.1086/158589

\bibitem[Johnstone et al.(2015)]{2015A&A...577A..28J} Johnstone, C.~P., G{\"u}del, M., Brott, I., et al.\ 2015, \aap, 577, A28. doi:10.1051/0004-6361/201425301


\bibitem[Loyd et al.(2022)]{2022ApJ...936..170L} Loyd, R.~O.~P., Mason, J.~P., Jin, M., et al.\ 2022, \apj, 936, 170. doi:10.3847/1538-4357/ac80c1

\bibitem[McMullin et al.(2007)]{2007ASPC..376..127M} McMullin, J.~P., Waters, B., Schiebel, D., et al.\ 2007, Astronomical Data Analysis Software and Systems XVI, 376, 127

\bibitem[Metcalfe et al.(2013)]{2013ApJ...763L..26M} Metcalfe, T.~S., Buccino, A.~P., Brown, B.~P., et al.\ 2013, \apjl, 763, L26. doi:10.1088/2041-8205/763/2/L26

\bibitem[Montes-Doria et al.(2022)]{2022MNRAS.509.1892M} Montes-Doria, D., Gonz{\'a}lez, R.~F., Cant{\'o}, J., et al.\ 2022, \mnras, 509, 1892. doi:10.1093/mnras/stab3085

\bibitem[Nita et al.(2002)]{2002ApJ...570..423N} Nita, G.~M., Gary, D.~E., Lanzerotti, L.~J., et al.\ 2002, \apj, 570, 423. doi:10.1086/339577

\bibitem[Reidemeister et al.(2011)]{2011A&A...527A..57R} Reidemeister, M., Krivov, A.~V., Stark, C.~C., et al.\ 2011, \aap, 527, A57. doi:10.1051/0004-6361/201015328

\bibitem[Rodriguez et al.(1993)]{1993RMxAA..25...23R} Rodriguez, L.~F., Marti, J., Canto, J., et al.\ 1993, \rmxaa, 25, 2

\bibitem[Rodr{\'\i}guez et al.(2019)]{Rodriguez19} Rodr{\'\i}guez, L.~F., Lizano, S., Loinard, L., et al.\ 2019, \apj, 871, 172. doi:10.3847/1538-4357/aaf9a6

\bibitem[Wang et al.(2020)]{2020RAA....20..178W} Wang, L., Liu, S.-M., \& Ning, Z.-J.\ 2020, Research in Astronomy and Astrophysics, 20, 178. doi:10.1088/1674-4527/20/11/178

\bibitem[Wild et al.(1963)]{1963ARA&A...1..291W} Wild, J.~P., Smerd, S.~F., \& Weiss, A.~A.\ 1963, \araa, 1, 291. doi:10.1146/annurev.aa.01.090163.001451

\bibitem[Wood et al.(2002)]{2002ApJ...574..412W} Wood, B.~E., M{\"u}ller, H.-R., Zank, G.~P., et al.\ 2002, \apj, 574, 412. doi:10.1086/340797

\end{thebibliography}
\end{document}